\documentstyle[preprint,aps]{revtex}
\begin{document}
\title{Spin-dependent transmission of holes through
periodically\\ modulated diluted magnetic semiconductor
waveguides}
\author{X. F. Wang and P. Vasilopoulos}
\address{Concordia University, Department of Physics,\\
Montr\'{e}al, H3G 1M8, Canada}
\address{}
\address{\mbox{}}
\address{\parbox{14cm}{\rm \mbox{}\mbox{}\mbox{}
We study
spin transport of holes through stubless or stubbed
waveguides modulated periodically by diluted magnetic
semiconductor (DMS) sections of width $b_1$.
Injected holes of {\it up (down)} spin feel a
{\it periodically modulated {\it barrier (well)}
potential} in the DMS sections and have different
transmission ($T$) coefficients. $T$ oscillates with
$b_1$ for spin-down and decreases fast for spin-up
holes while the
relative polarization $P_r$ depends
nearly periodically on the stub height. Using asymmetric
stubs leads to a nearly {\it square-wave} pattern in $T$
and to wide plateaus in $P_r$. $T$ oscillates 
with the
length between the DMS sections. 
With two DMS
sections per unit, $T$ shows periodically wide gaps for
spin-down holes when a DMS width  is varied.
The results can be used to create efficient spin filters. }}
\maketitle

\vspace{5mm}
\newpage
In recent years spin transport has attracted considerable
attention as it offers a possibility for a new type of
transistor \cite{dat}, quantum computation and quantum logic
\cite{tod}.
A significant part of the work has concentrated on spin-polarized electronic
transport through diluted magnetic semiconductors (DMS) \cite{tod}
despite the fact that the reported experimental spin
polarizations are very low, about 1\%, and make the
results controversial and attributable to extraneous effects. Experimental work \cite{mat}
showed that DMS materials $Ga_{1-x}Mn_{x}As$ can be in the  metallic
ferromagnetic phase with heavy holes as carriers for
$0.03\leq x\leq 0.05$.

In previous work \cite{wan} we demonstrated that the
electronic transmission through periodically stubbed
waveguides, in the presence of spin-orbit interaction,
shows a rich structure and a spin-transistor behavior
as a function of the stub parameters. Here we demonstrate
similar effects in the transmission of heavy holes
through waveguides made of many identical units each of which consists of a
nonmagnetic part, A, followed by a DMS one, B, as shown
in Fig. 1 (a). Stubs can be attached to either part but
it is advantageous to attach them to part B.

The spin-structure for holes is obtained \cite{gha}
self-consistently in reciprocal space. The hole interaction
with magnetic impurities is described by the potential

\begin{equation}
U_{mag}({\bf {r})=
-}I{\bf \sum_{i=1}^{N_{i}}{s}({r}).{S}({R_{i}}) \delta
({r}-{R_{i}}),}
\label{umag}
\end{equation}
where $I$ is the $p-d$ exchange coupling constant,
${\bf {R_{i}}}$ denotes the positions of the $N_{i}$
impurities $Mn$ , uniformly distributed in the DMS layers,
${\bf{S}({R_{i}})}$ is the spin of the
impurity, and
${\bf{s}}$ the spin of the hole. We assume the magnetization
of each layer to be oriented along a single direction, each
layer being in its ferromagnetic phase. Thus, the spin of the
hole is well defined in this direction, being parallel (down) or
antiparallel (up) to it. Integrating $U_{mag}({\bf{r}})$
over z and x gives the effective
potential
$V_{mag}(y)$ in terms of the average magnetization
$<{\bf {M}>_{j}}$:
\begin{equation}
V_{mag}^{\sigma}(y)=V_0\sigma\sum_{j}<M>_{j}g_{j}(y),
\label{effecmag}
\end{equation}
where $g_{j}(y)=1$ if $y$ lies inside the $j$-layer, and
$g_{j}(y)=0$ otherwise; $\sigma= \pm 1$ for spin-up and spin-down
holes, respectively, and $V_0$ is a sample-dependent parameter for
the strength of the potential. The total potential felt by a hole
is $U^{\sigma}_{eff}(y) =[U_{c}(y)+V_{mag}^{\sigma}(y)+U_{h}(y)]
$, where $U_c(y)$ is the confining potential, and $U_h(y)$ the
Hatree hole-hole interaction potential. Setting
$E_k=h^2k^{2}/2m^*$ the hole Hamiltonian in the DMS section
becomes
\[H=E_k+U^{\sigma}_{eff}(y)\equiv
\left[
\begin{array}{cc}
E_k+U^{+}_{eff}(y) & 0 \\
0 & E_k+U^{-}_{eff}(y)
\end{array}\right]
\]
The direction of the spin polarization
depends on that of $<{\bf {M}>_{j}}$.
We assume the latter is along the waveguide (y axis)
or along the stub (x axis).
Generally, $U^{\sigma}_{eff}(y)$ is not constant inside
the DMS layer and
depends on parameters such as the Mn density, the layer
size, the temperature, etc. For simplicity we assume that
$U^{\sigma}_{eff}(y)$ is constant and $U^{\pm}_{eff}=\pm V$.
A $y$-dependent potential
can be treated by considering a DMS layer as a seriers of equal-width 'flat'
potential layers of unequal height. Even for high hole densities, the resulting
wave functions can be well approximated \cite{gha} by those of a square well.

The subject of this paper is a transistor-like modulation of
a spin current.
Motivated by previous results on electronic
stub tuners
\cite{wan,tak1} 
we consider ballistic spin transport of holes through
waveguides with double
stubs attached to them as in Fig. 1 (a) or without stubs.

Since the Hamiltonian is diagonal in the spin index, we can
consider the propagation of spin-up and spin-down holes inside the
structure independently and denote the wave functions by $\phi^+$
and $\phi^-$, repectively. In each region of Fig. 1 (c) we have
$\phi _{n}^{\sigma}(x)=\sqrt{2/w}\sin (n\pi (x+w/2)/w)$, where $w$
is the width of the region along $x$. Including spin we can write
the eigenfunction $\phi _{1}^{\sigma }$ of energy $E$ in region I
as

\begin{equation}
\phi _{1}^{\sigma }=\sum_{m}\{a_{1m}^{+}e^{i\beta _{m}^{\sigma}y}
+b_{1m}^{+}e^{-i\beta _{m}^{\sigma }y}\}\sin [c_{m}(x+c/2)]
\label{wavf}
\end{equation}
Here $c_{m}=m\pi /c$ and $\beta _{m}^{\sigma }=[2m^{\ast }
(E-U_{eff}^{\sigma})-c_{m}^{2}]^{1/2}$.
In region III $\phi _{2}$ is given by Eq. (\ref{wavf})
with the changes $1m\rightarrow 2m$,
$U_I^{\sigma}\rightarrow
U_{III}^{\sigma}$, and $y\rightarrow y-b_1$.
In the stub region II, Eq. (\ref{wavf}) remains valid with the
changes
$c\rightarrow h$,
$U_I^{\sigma}\rightarrow U_{II}^{\sigma}$,
and $x+c/2\rightarrow x+h/2-d$.

Matching the wave function and its derivative at $y=0$ and
$y=b_1$ connects the incident waves (to the left of region I)
with the outgoing ones (to the right of region III) with a
spin-dependent transfer matrix $\hat{M}$

\begin{equation}
\left( \begin{array}{c}a_{in}^{\sigma } \\ b_{in}^{\sigma }
\end{array}\right) =\hat{M}^{\sigma }\left(
\begin{array}{c}a_{out}^{\sigma } \\ b_{out}^{\sigma }
\end{array}\right).
\end{equation}

In DMS layers the hole momentum $\beta^{\sigma}$ is strongly
spin-dependent; this results in a phase difference between
the spin-up and
spin-down wave functions and in the spin-dependence of the
transfer matrix
$\hat{M}^{\sigma}$. If we inject {\it unpolarized} holes to
the left of the
structure, we can obtain spin {\it polarized} outgoing holes
to its right
because the transfer matrix $\hat{M}^{\sigma }$ and the
transmission
$T^{\sigma }$ depend strongly on the spin $\sigma $ of
the holes.
Here are the results.

In Fig. 2 we plot $T$ vs the stub width $b_1$ for the parameters
given in the caption. As shown, 
for a simple
unit the spin-up transmission $T^+$ decreases very fast whereas
the spin-down one $T^-$ oscillates. This fast decrease can
be used to filter out the spin-up holes. This result agrees
qualitatively with that for an unconfined single DMS layer
in the presence of a normal magnetic field
\cite{cha}. For a composite unit ($b_{1}\neq b_{2}$)
the result is similar. An interesting feature is shown
in the inset. $T^\pm$ 
oscillates with the length between the DMS segments, i.e., it shows
{\it longitudinal} resonances as that of spinless electrons
\cite{kir}. 

The values of $l$ at which $T^+$ maxima
occur, $l_m$, in Fig. 2
can be utilized to maximize the gaps in
the transmission. In Fig. 3 we show $T$ vs $b_1$ for a
waveguide with 40 composite units and two different temperatures.
Any $l_m$ value leads to a maximum gap as dephasing of the
wave
between the DMS and GaAs sections is maximal.

So far we considered waveguides without stubs. If we attach stubs,
we obtain several interesting results. The first of them is shown
in Fig. 4 where we plot, for {\it symmetric} stubs, the relative
output polarization $P_{r}=(J_{u}-J_{d})/(J_{u}+J_{d})$ vs the
stub height $h$, with $J_{u}$ ($J_{d}$) being the up (down)
current. As $N$ increases the gaps become wider and deeper and
reach a limit for $N=6$. Notice how $P$ switches, nearly
periodically, between +1 and -1, as $h$ is changed. This behavior
can be understood by inspection of the inset where we plot
separately the currents $J_u$ and $J_d$ vs $h$. Due to the large
effective Zeeman splitting the dependence of $T^+\propto J_u$ and
$T^-\propto J_d$ on $h$ is different and translates directly to
that of $P_{r}$.

Another interesting result is shown
in Fig. 5 for {\it asymmetric} stubs where
$P_r$ is plotted vs the
asymmetry parameter $d$, cf. Fig. 1 (c). Notice the wide ranges of
$d$ in which $P_r$ takes the values -1, +1, and 0.
The nearly {\it square-wave} pattern of $T$ shown in
the inset is similar to that for electrons with spin
neglected \cite{tak1} or considered in the presence of spin-orbit
interaction \cite{wan}.

A qualitative understanding of the results shown in Figs.
2-5 is easily reached if we combine the basic idea of a
stub tuner \cite{dat2} and its refinements \cite{tak1}
with the effective, spatially modulated potential of
Fig. 1 (b) which makes {\it holes of up (down) spin
"see" a barrier (well)} when propagating through the DMS
sections. In a stub tuner waves reflected from the walls
of the stub,
where the wave function vanishes, may interfere
constructively or destructively with those in the main waveguide
and result, respectively, in an increase or decrease of the
transmission. Refining this idea showed \cite{tak1} that using
{\it asymmetric} double stubs the transmission of {\it spinless}
electrons could be blocked completely. Combining several stubs
leads to a nearly {\it square-wave} transmission as a function
of the asymmetry parameter $d$.
The same idea applies to the holes we consider here
and the transmission shown in the inset of Fig. 5 is simply the
result of this behavior. In addition, the large effective
Zeeman splitting between the
spin-up and spin-down holes readily explains the oscillations
of $T^-$ in Figs. 2-3, since spin-down holes "see" a well
whereas spin-up holes "see" a barrier which results
in $T^+\to 0$ upon increasing its width. The behavior of
$P_r$ in Figs. 4-5 stems directly 
from that of
$J_u$ and $J_d$ which have a different dependence
on $h$ or $d$ due to the large Zeeman splitting. As for the
{\it longitudinal} resonances of $T^\pm$ in Fig. 2 , they
have the same origin as those for spinless electrons \cite{kir}:
in essence they result from matching of the
phase of the wave between the GaAs and DMS sections. In
contrast, absence of this matching leads to the large
gaps of $T^-$ in Fig. 3. Finally, with regard to the
temperature dependence of some of the results, the
reduction (and rounding off) of the transmission
peaks and of the gaps in Fig. 3 is expected and
in line with that for electrons \cite{wan,tak1};
it can be offset by increasing
the ratio of the Fermi to the thermal energy.
An important question concerns the influence of the stub shape on the transmission output.
But as in electronic stub tuners \cite{wan,tak1},
we have verified that here too changing the stub shape does not change
the transmission qualitatively; it only alters its
period when plotted, e.g., vs $h$, and several units are combined.

The results presented so far are valid when only a
single mode propagates in the main waveguide. If more
modes propagate, the transmission pattern becomes more complex
but it is still possible to have a periodic output, as in Fig. 2, if
$b_1$ is short enough that only a single mode penetrates in the stub
\cite{tak1}. Details will be given elsewhere.

The 
DMS devices, on which the observability of the results relies, could be  fabricated
using the recently developed low-temperature  MBE 
technique \cite{mat}  to grow a  superlattice which
could be etched perpendicularly to produce a surface superlattice.
Then  patterned gates can be deposited on its surface
to control the shape of
the stubs. 

In summary, we combined the spin-dependent transmission
through a DMS section with the basic physics of a stub
tuner and applied it to hole transport through a stubless
or stubbed waveguide. We showed that the large effective
Zeeman splitting results in a spin-dependent transmission.
The transmission of the spin-up holes can be blocked
whereas that of the spin-down ones oscillates with the
DMS width showing wide gaps. More important, in stubbed
waveguides the relative polarization varies nearly periodically
with the stub height, switching from -1 to +1, while the
transmission shows a nearly {\it square-wave} pattern
upon using {\it asymmetric} stubs.
The results should lead to the creation of efficient spin filters.

 We thank Dr. I.C. da Cunha Lima for helpful discussions.
Our work was supported by the Canadian NSERC Grant No.
OGP0121756.

\begin{figure}[tbp]
\caption{\label{f1}(a) A periodically stubbed waveguide
(A=GaAs, B=DMS).
(b) The effective potential V along the growth axis for
spin-up (solid curve)
and spin-down (dotted curve) holes. (c) A waveguide portion
with two stubs
(shaded areas). For a simple unit we have $b_{1}=b_{2}$, for
a composite one
$b_{1}\neq b_{2}$.
The midpoints of $h$ and $c$ determine the asymmetry parameter
$d$.}
\end{figure}
\begin{figure}[tbp]
\caption{\label{f2} Transmission $T$ as a function of the stub
width $b_1$
for a simple unit (solid and dotted curves)
and for a composite one (dashed and
dash-dotted curves, $b_{2}=50$ \AA with $l=207.5$ \AA.)
The fast decreasing curves are for spins up ($T^+$),
the oscillating ones for spins down ($T^-$).
The inset shows $T^+$ (solid curve) and $T^-$
(dotted curve) vs $l$ (in units of 100 \AA )
of a stubless ($h=a=250$ \AA ) waveguide
with $b_{1}=b_{2}=50$ \AA, $V=3$ meV, and
$E_{F}=4.48$ meV.}
 \end{figure}
\begin{figure}[tbp]
\caption{ \label{f3} Transmission $T$ vs 
the 
width
$b_{1}$
in a {\it stubless}  ($h=a=150$ \AA ) waveguide of 40 composite
units
($b_{2}=30$ \AA , $l=142$ \AA, $V=16.5$ meV,
$E_{F}=12.5$ meV).
The solid (dashed) curves show $T^\mp$ 
at
the indicated temperatures.}
\end{figure}

\begin{figure}[tbp]
\caption{ \label{f4} The relative output polarization $P_{r}$ vs
the stub height $h$ for $V=0.3$ meV, $E_{F}=4.48$ meV, and
temperature $T=0.1$ K. $N$ is the number of simple units
($b_{1}=b_{2}=50$ \AA ). The inset shows $T^+\propto J_u$
 (solid curve)
and $T^-\propto J_d$ (dotted curve) vs $h$.}
\end{figure}
\begin{figure}[tbp]
\caption{\label{f5} The relative output polarization
$P_{r}$ vs the asymmetry parameter $d$ for a stubbed
waveguide of 5 simple units
($b_{1}=b_{2}=150$ \AA, $l=207.5$ \AA, $h=1014$ \AA )
at temperature $T=0.1$ K with
$V =0.3$ meV and
$E_{F}=4.48$ meV. The inset shows $T$ versus $d$
with $b_{1}=b_{2}=50$ \AA\
and $l=207.5$ \AA.
The solid (dashed)
curves are for spin-down (spin-up) holes. }
\end{figure}

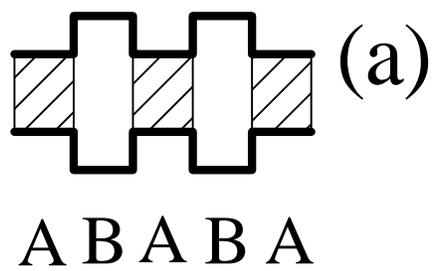
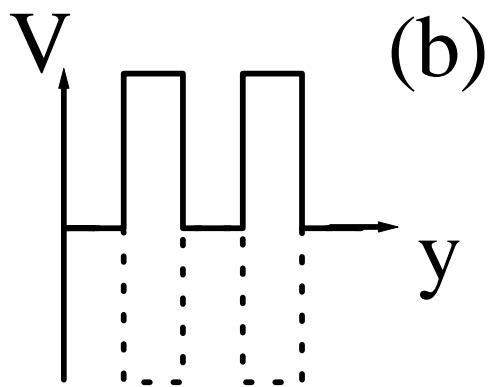
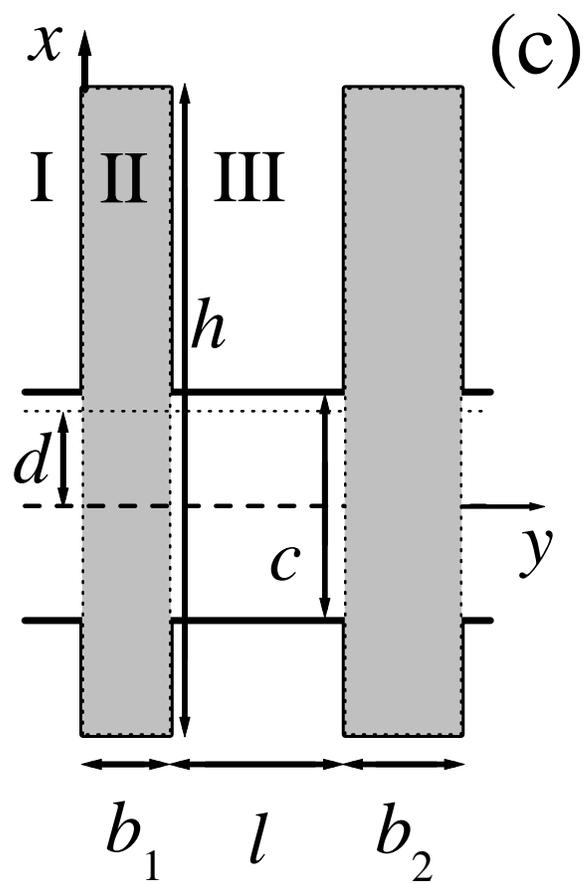
\end{document}